\documentclass[twocolumn,preprintnumbers,aps,prb,superscriptaddress]{revtex4}
\usepackage{amsmath}
\usepackage{graphicx}
\usepackage{textcase}
\usepackage{bm}

\begin{document}

\title{Enhanced Pairing in the ``Checkerboard'' Hubbard Ladder}
\author{George Karakonstantakis}
\affiliation{Department of Physics, Stanford University, Stanford,
CA 94305, USA}
\author{Erez Berg}
\affiliation{Department of Physics, Harvard University, Cambridge,
MA 02138, USA}
\author{Steven R. White}
\affiliation{Department of Physics and Astronomy, University of California, Irvine, CA
92717, USA}
\author{Steven A. Kivelson}
\affiliation{Department of Physics, Stanford University, Stanford,
CA 94305, USA}
\date{\today}

\begin{abstract}
We study signatures of superconductivity in a 2--leg
``checkerboard'' Hubbard ladder model, defined as a
one--dimensional (period 2) array of square plaquettes with an
intra-plaquette hopping $t$ and inter-plaquette hopping $t'$,
using the density matrix renormalization group method. The highest
pairing scale (characterized by the spin gap or the pair binding
energy, extrapolated to the thermodynamic limit) is found for
doping levels close to half filling, $U\approx 6t$ and $t'/t
\approx 0.6$. Other forms of modulated hopping parameters, with
periods of either 1 or 3 lattice constants, are also found to
enhance pairing relative to the uniform two--leg ladder, although
to a lesser degree. A calculation of the phase stiffness of the
ladder reveals that in the regime with the strongest pairing, the
energy scale associated with phase ordering is comparable to the
pairing scale.
\end{abstract}

\maketitle

\section{Introduction\label{sec:Intro}}

The much debated theoretical issues related to the ``mechanism'' (\textit{%
i.e.} microscopic origin) of high temperature superconductivity
are often ill-defined. One
related question to which unambiguous answers
are possible is: For a given class of models,
what values of the parameters are optimal for superconductivity? Of course,
if one can make predictions about models, 
the same insights \emph{might} provide guidance in the search for materials
with improved superconducting properties. Two specific questions we would
like to address are: 1) In the case in which superconductivity arises
directly from the repulsive interactions between electrons, how strong (in
units of the bandwidth) are the optimal interactions for superconductivity?
2) Is there an ``optimal inhomogeneity'' for superconductivity \cite%
{Steve&Fradkin}, in the sense of a complex (but still periodic)
electronic structure with multiple orbitals per unit cell? An
obvious difficulty with this program is that, in most cases, we do
not know how to compute the transition temperature of the relevant
models in a controlled manner, so as to test the predictions of
theory.

In this context, we use density matrix renormalization group
(DMRG)\cite {DMRGoriginal} to numerically compute the
superconducting correlations of the two-leg Hubbard ladder
(extrapolated to infinite length) as a function of the strength of
the Hubbard interaction, $U$, and for various periodic patterns of
the hopping matrix elements. The 1D character of the system
studied is what permits us to obtain an accurate solution of this
problem. However, the same 1D character implies that no non-zero
critical temperature is possible, so in assessing the optimal
conditions for superconductivity, we are forced to use other
energy scales in the problem, especially the spin-gap, $\Delta E
_{s}$, the pair-binding energy, $\Delta E _{p}$, and the
superfluid helicity modulus, $\rho _{c}$. We find that: 1) The
optimal value of $U$ is generally $U\approx 6t$ where $6t$ is the
total bandwidth of the uniform ladder. This result agrees with
previous studies\cite{2LegHubbard,Noack-1997} of various ladder
systems. It is also consistent with inferences made on the basis
of exact diagonalization\cite{Hong2} and dynamical cluster quantum
Monte-Carlo\cite{Maier-2006} studies of the 2D Hubbard model,
where $U\approx 8t$ (\textit{i.e.} the 2D bandwidth) was found to
be optimal. 2) For the checkerboard pattern with 4 sites per unit
cell shown in Fig. 1b, the optimal conditions occur for an
intermediate degree of inhomogeneity, $t^{\prime }/t\sim 0.6-0.7$,
where $t$ is the hopping matrix within a square and $t^{\prime}$
is the hopping matrix between squares. This tends to corroborate
inferences made previously on the basis of exact diagonalization
studies\cite{Hong2} of the 2D \textquotedblleft
checkerboard-Hubbard model.\textquotedblright\
3) 
A qualitatively similar enhancement of superconductivity is observed 
 for the 
other periodic versions of the model
with 2 
or 6 sites per unit
cell 
shown in Figs. 1a and 1c, respectively, although in these cases the magnitude of the effect is smaller and the optimal condition occurs with values of $t^{\prime }/t$ closer to 1.

\begin{figure}[b!]
\includegraphics[width=0.5\textwidth]{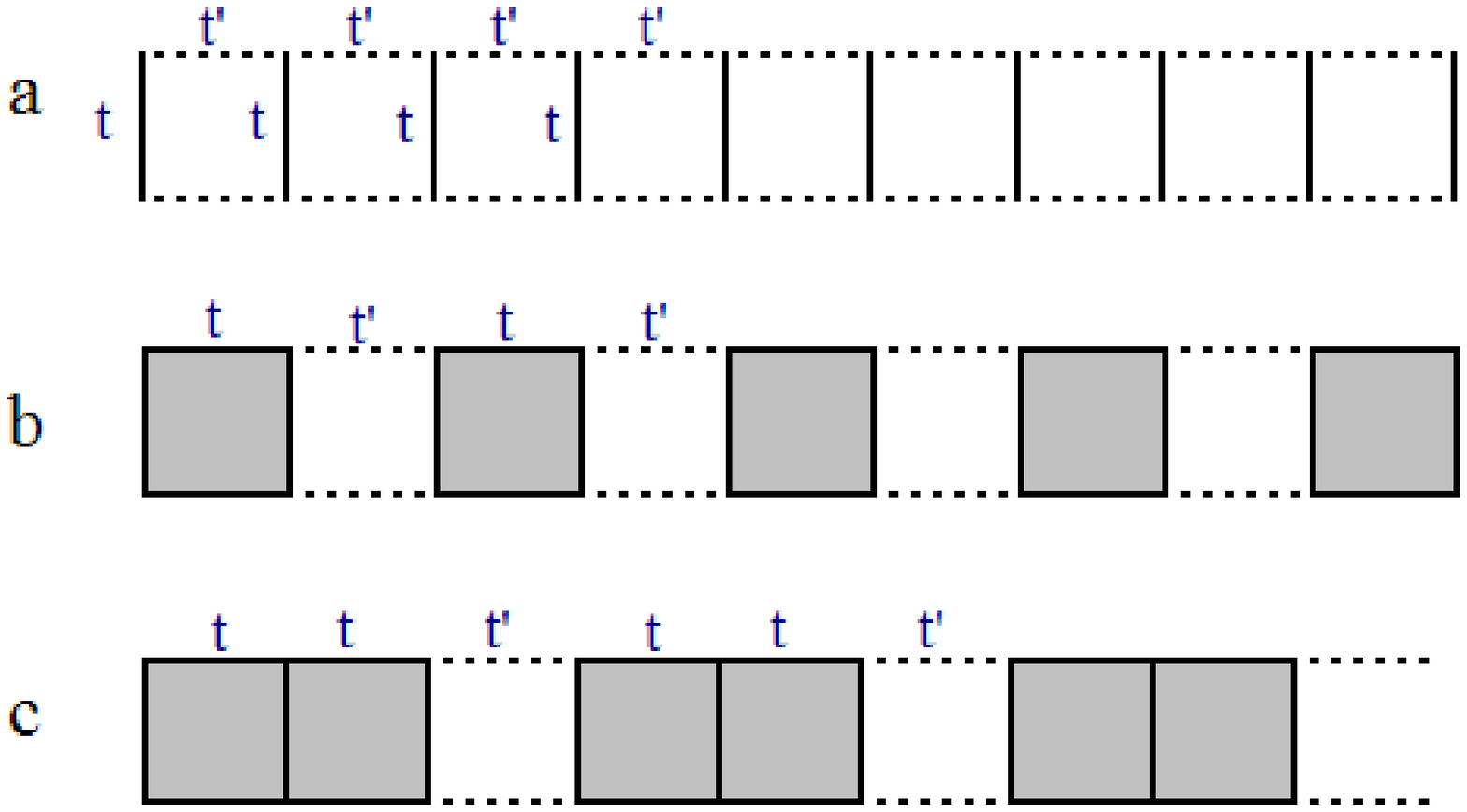}\vspace{-0.4in}
\caption{ Schematic representation of the ``inhomogeneous''
Hubbard ladders considered in the present paper: a) The period one
``dimer'' ladder; b) the period two ``checkerboard'' ladder; c)
the period three ladder. As discussed below Eq.
\protect\ref{eq:checker}, the solid and dashed lines represent,
respectively, hopping matrix elements $t$ and $t^\prime$.}
\label{fig:ladder}
\end{figure}

The observation that certain patterns of spatial symmetry breaking
can coexist with superconductivity (or even strongly enhance it),
while others do not, is also reminiscent of recent results
obtained using DMRG\cite{White-2009} and the dynamic cluster
approximation\cite{Maier-2010}. In the first of these
calculations, the inhomogeneity (in the form of stripes) occurs
spontaneously, while in the second it is imposed externally. As we
were completing this work, a contractor renormalization (CORE)
study of the checkerboard Hubbard model in a 2D geometry was
presented in Ref.~ \onlinecite{Baruch-2010}, extending earlier
CORE results for the uniform 2D Hubbard model\cite{Altman-2002}.
Finite size effects were found to be large for $t^{\prime}\gtrsim
0.8t$, but in the smaller $t^{\prime}$ regime, where these effects
are relatively small, the results of this new study are completely
consistent with those of the earlier exact diagonalization
studies\cite{Hong2}, and lead to conclusions concerning the
optimal conditions for superconductivity that are similar to those
obtained in the present ladder study. The CORE method was also
used to study ladders, albeit considerably shorter than those
studied in the present paper, and again the results obtained are
fully consistent with the present results.

\section{The model\label{sec:model}}

We consider the repulsive $U$ Hubbard model defined on a (spatially
modulated) two-leg ladder
%
\begin{align}
\mathcal{H}=& -\sum_{j,\lambda ,\sigma }(t_{j,j+1}c_{j,\lambda ,\sigma
}^{\dagger }c_{j+1,\lambda ,\sigma }+h.c.)  \notag  \label{eq:checker} \\
& -t\sum_{j,\sigma }(c_{j,1,\sigma }^{\dagger }c_{j,2,\sigma
}+h.c.)+U\sum_{j,\lambda }n_{j,\lambda ,\uparrow }n_{j,\lambda ,\downarrow }
\end{align}%
Here $c_{j,\lambda ,\sigma }^{\dagger }$ 
creates 
an electron 
on rung $j=1,\ldots \ L-1$ 
of chain $\lambda =1,\ 2$ with spin polarization $\sigma =\pm $, $L$ is the
length of the ladder,
$U>0$ is 
the repulsion between two electrons on the same site, the density operator
is $n_{j,\lambda ,\sigma }=c_{j,\lambda ,\sigma }^{\dagger }c_{j,\lambda
,\sigma }$, and $n=(2L)^{-1}\sum_{j,\lambda ,\sigma }\langle n_{j,\lambda
,\sigma }\rangle $ is the mean number of electrons per site.
The much studied homogeneous Hubbard ladder 
corresponds to the case $t_{j,j+1}=t^{\prime }$ for all $j$, although it is
worth noting that for $t^{\prime }\ll t$, this model can also be viewed as a
coupled array of Hubbard-dimers.
The \textquotedblleft dimer ladder\textquotedblright\ is shown in Fig. \ref%
{fig:ladder}a. 
The \textquotedblleft 
checkerboard ladder\textquotedblright\ in Fig. \ref{fig:ladder}b has $%
t_{2j,2j+1}=t$ and $t_{2j+1,2j+2}=t^{\prime }<t$. The \textquotedblleft
period three\textquotedblright\ ladder in Fig. \ref{fig:ladder}c has $t_{{%
3j,3j+1}}=t_{3j+1,3j+2}=t$ and $t_{3j+2,3j+3}=t^{\prime }<t$.
The thermodynamic limit is accessed by computing quantities for various
lengths, and then using finite size scaling analysis to extrapolate to $%
1/L\rightarrow 0$.


\section{Effective Field-theory}

The uniform two-leg Hubbard ladder with $n\neq 1$ but still not too far from
$n=1$, is well known, on the basis of weak coupling RG\cite{Ballents&Fisher}%
, bosonzation\cite{Schultz}, and DMRG\cite{2LegHubbard} approaches, to be in
a Luther-Emery phase characterized at low energies by a spin-gap, $\Delta E
_{s}$ (defined in Eq. \ref{eq:spingap}, below) and a single, gapless
acoustic \textquotedblleft charge\textquotedblright\ mode which propagates
with speed $v_{c}$, and whose long-range (power-law) correlations are
determined by a single Luttinger parameter, $K_{c}$. The Luther-Emery liquid
can be thought of as a 1D version of a superconducting state in the sense
that it has a non-vanishing superfluid stiffness (see Eq. \ref{eq:stiff},
below), and, for $K_{c}>1/2$ and $T\ll \Delta E _{s}$, it has a divergent
superconducting susceptibility,
\begin{equation}
\chi \sim \chi _{0}\left( \frac{v_{c}}{aT}\right) ^{(2-1/K_{c})},
\label{chi}
\end{equation}%
where $v_{c}$ is the charge velocity and $a$ is a lattice
constant. In the single chain realization of a Luther-Emery
liquid, 
\begin{equation}
\chi _{0}=\left( \frac{a}{v_{c}}\right) \left( \frac{a\Delta E _{s}}{v_{s}}%
\right) =\left( \frac{a}{v_{c}}\right) \left( \frac{a}{\xi
_{s}}\right) ,
\end{equation}%
where $\xi _{s}= v_{s}/\Delta E _{s}$ is the spin-correlation length and $%
v_{s}$ is the spin-velocity. For a multicomponent system, the corresponding
expression for $\chi _{0}$ is somewhat more complicated, as there may be
multiple scales (\textit{e.g.} multiple spin-gaps) associated with the
gapped modes. However, $\chi _{0}$ remains a monotonic, approximately
linearly increasing function of $\Delta E _{s}$.

Perhaps not surprisingly, we will see that the inhomogeneous Hubbard ladders
we have studied are also Luther-Emery liquids with $K_c > 1/2$. Thus, in
addressing the ``mechanism of superconductivity,'' the primary purpose of
our DMRG calculations is to determine the dependence of $v_{c}$, $K_{c}$, $%
\Delta E _{s}$ and $\xi_s$ on microscopic parameters.



The pair binding energy $\Delta E_p$ corresponds to creating two
spatially-separated spin-1/2 quasiparticles. Since the spins for
these quasiparticles can either add to $S=0$ or $1$, we must have
$\Delta E_s \le \Delta E_p$. If the residual interactions between
quasiparticles are repulsive, we expect $\Delta E_s = \Delta E_p$.
Conversely, if the interactions between quasiparticles are
attractive, a neutral spin-1 ``exciton'' is formed, which has
lower energy than two far-separated quasi-particles, and hence
$\Delta E_s < \Delta E_p$. The latter behavior has been found
previously in DMRG calculations on the uniform $t-J$
ladder\cite{Poilblanc-2000}.

\section{Energy scales}

\label{sec:scales} 

The spin-gap, $\Delta E _{s}$, is the difference between the
ground-state energies of the
system with spin $S=1$ and $S=0$: 
\begin{equation}
\Delta E _{s}\equiv \mathcal{E}_{0}(S=1,2N)-\mathcal{E}_{0}(S=0,2N),
\label{eq:spingap}
\end{equation}%
where $\mathcal{E}_{0}(S,N)$ is the spin $S$ ground-state energy
of the $N$ electron system.

Similarly, the pair-binding energy, $\Delta E _{p}$, is defined as
\begin{equation}
\Delta E _{p}=2\mathcal{E}_{0}(\frac{1}{2},2N+1)-\mathcal{E}_{0}(0,2N)-\mathcal{E}%
_{0}(0,2N+2) .  \label{eq:pair}
\end{equation}%
Were we computing these quantities in a BCS superconductor, then
in the thermoydnamic limit, both these energies would be equal to
twice
the minimum gap $\Delta _{\mathrm{min}}$

\begin{equation*}
\lim_{L\rightarrow \infty }\Delta E _{s}=\lim_{L\rightarrow \infty }\Delta
E_{p}=2\Delta _{\mathrm{min}} .
\end{equation*}
%
Thus, it is intuitively reasonable to associate these energy scales with a
mean-field estimate of the superconducting critical temperature, $%
T_{c}^{MF}\sim \Delta E_s /4$. Of course,
since the ladder is a 1D system, the actual $T_{c}=0$.




While it may be reasonable to interpret $\Delta E _{s}$ and/or $\Delta E
_{p} $ as measures of a pairing scale in the problem, in order to address
the growth of superconducting correlations it is ultimately necessary to
consider the helicity modulus, which governs the energetics of
superconducting phase fluctuations:
\begin{equation}
\rho _{c}=\frac{v_{c}K_{c}}{2\pi }\equiv \lim_{L\rightarrow \infty }\left[L%
\frac{\partial ^{2}\mathcal{E}_{0}}{\partial \phi ^{2}}\Big|_{\phi =0}\ %
\right]  \label{eq:stiff}
\end{equation}%
where, in this case, the ground-state energy is computed in the presence of
pair-fields applied to the two ends of the system with a relative phase
twist $\phi $.
%

In 2D, the relative importance of phase and pair-breaking fluctuations can
be assessed\cite{EmeryKivelson} by considering the ratio of the phase
stiffness (which has units of energy) to the pairing gap. However, in 1D, $%
\rho _{c}$ has units of a velocity, so defining an energy scale,
$\Delta E _{\theta }$, characteristic of the phase fluctuations
requires introducing a
length scale in the problem. 
The only emergent length scale is
$\xi _{s}$, 
in terms of which we define
\begin{equation}
\Delta E _{\theta }\equiv \pi\rho _{c}/\xi _{s}\equiv R\ \Delta
E_s. \label{eq:phase}
\end{equation}
Here $R\equiv \Delta E_\theta / \Delta E_s$ 
is the dimensionless ratio of the phase ordering and pairing scales.

To appreciate the significance of this ratio, consider its value
for the attractive Hubbard chain in various limits. The 1D version
of a BCS limit, in which there is a single characteristic
energy/temperature scale, $\Delta_s \sim \exp[-\pi v_s/a|U| ]$, is
realized in the limit $|U| \ll 1$ where, up to corrections of
order $U/t$, $v_s = v_c$ and $K_c=1$, so $R=
v_cK_c/2v_s=1/2+\mathcal{O}(U/t)$, \textit{i.e.} both mesoscale
phase coherence and pairing correlations onset at a temperature of
the order of $T_{\mathrm{pair}}\sim \Delta E_s/4$. Conversely, $R
\to 0$ as $|U|/t \to \infty$; for large $U$, a spin pseudo-gap
opens when $T \sim T_{\mathrm{pair}} 
= |U|/2$, with a second crossover
from a largely incoherent paired state to a coherent Luther-Emery
liquid occurring at a temperature $T_{\mathrm{\theta}} \sim \Delta
E_{\theta} \propto t^2/|U|$, well below $T_{\mathrm{pair}}$. A
similar dichotomy exists in the two-leg repulsive $U$ Hubbard
ladder, where $R \to 0$ as the doped hole concentration, $x \to
0$, while $R \sim 1$ at larger values of $x$ where the spin-gap is
significantly suppressed relative to its value at $x=0$. In the
small $x$ case, the doped holes can be thought of as
a dilute gas of charge $2e$ bosons at temperatures small compared to $T_{%
\mathrm{pair}}$, but the phase coherence scale is much smaller and vanishes
as $x\to 0$.%

With these examples in mind, we identify the case $R\sim 1$ with the 1D
version of
the ``BCS-like limit'' in which there is a single crossover temperature $T_{%
\mathrm{pair}}$ which separates the \textquotedblleft
normal\textquotedblright\ (multicomponent Luttinger liquid) high temperature
regime from the low temperature regime in which substantial mesoscale
superconducting order has developed. Conversely, if
$R \ll 1$, two distinct crossover scales characterize the evolution from the
normal state: a first, high temperature crossover, $T_{\mathrm{pair}}$,
characterized by the opening of a spin pseudo-gap, and a lower crossover
temperature, $T_{\theta }\sim \Delta E _{\theta }/4$, which can be viewed as
the scale at which the liquid of 
bosonic pairs begin to exhibit substantial local phase coherence.

The most direct and efficient way to compute $\xi _{s}$ from DMRG is to
apply a staggered Zeeman field to one end of the ladder, $j=0$ (thus locally
breaking spin-rotational symmetry) and then measure the decay of the
magnetization as a function of distance down the ladder. In a spin-gapped
phase, we expect 
\begin{eqnarray}
M(j) &=&\sum_{\sigma }\sigma \langle \lbrack c_{j,1,\sigma }^{\dagger
}c_{j,1,\sigma }-c_{j,2,\sigma }^{\dagger }c_{j,2,\sigma }]\rangle  \notag \\
&\sim &\cos [Qj+\phi _{0}]\ \exp [-|j|a/\xi _{s}].  \label{Mi}
\end{eqnarray}%
In the limit of an asymptotically small spin-gap, $Q=2k_{F}$, but
for larger gaps it may depend not only on $n$ but on $U/t$ as
well. To be explicit, we therefore define the spin correlation
length as
\begin{equation}
\xi _{s}=\frac{\sum_{j}|j\ M(j)|}{\sum_{j}|M(j)|}.
\label{eq:spincorr}
\end{equation}%

It turns out that Eq. \ref{eq:stiff} is relatively difficult to
implement to obtain quantitatively reliable results for $\rho_c$
using DMRG. However, it
is possible \cite{Friedel} to compute $\rho _{c}$ by separately calculating $%
v_{c}$ and $v_{c}/K_{c}$ from quantities that are more straightforwardly
computed using DMRG.
From the bosonized field theory, we can identify the inverse
compressibility of the ladder with the ratio $\frac{\pi
v_{c}}{2K_{c}}$. In turn, in all circumstances relevant to the
present calculation\cite{Chang-2007}, the compressibility is
related to the energy to add or remove a singlet pair of electrons
from the ladder:
\begin{equation}
\frac{1}{\kappa }=\lim_{L\rightarrow \infty }L\frac{\mathcal{E}_{0}\left(
0,2N+2\right) +\mathcal{E}_{0}\left( 0,2N-2\right) -2\mathcal{E}_{0}\left(
0,2N\right) }{4},  \label{compressibility}
\end{equation}
An independent measurement of $v_{c}$ can be obtained by calculating also
the energy of the first excited state, $\mathcal{E}_{1}\left( S,N\right) $
according to
\begin{equation}
v_{c}=\lim_{L\rightarrow \infty }\frac{L}{\pi }\left[ \mathcal{E}_{1}\left(
0,2N\right) -\mathcal{E}_{0}\left( 0,2N\right) \right] .  \label{vc}
\end{equation}
We then compute the helicity modulus 
as
\begin{equation}
\rho _{c}=\frac{v_{c}^{2}\kappa }{4}\text{.}  \label{rhoc}
\end{equation}
Note that this procedure also gives us
\begin{equation}
K_{c}=\frac{\pi }{2}\kappa v_{c}.  \label{Kc}
\end{equation}
An alternative way to obtain $K_c$ is by measuring the
amplitude of the ``Friedel-like'' density oscillations which 
exhibit a power-law decay as a function of distance from the edge of the system. For long systems, the density near the center of a length $L$ ladder takes the
form\cite{Friedel}
\begin{equation}
\langle n_j \rangle \sim \frac{\cos[2\pi n
(j-L/2)]}{L^{K_c/2}}\text{.} \label{Kc_density}
\end{equation}
Therefore, by measuring the
amplitude of the density oscillations $A_{\mathrm{CDW}}$ vs. $L$
and plotting $\log(A_{\mathrm{CDW}})$ vs. $\log(L)$, $K_c$ can be
obtained. Whenever possible, we have calculated $K_c$ using both
Eq. {\ref{Kc}} and Eq. \ref{Kc_density}, and found that the two
values agree with each other to within about $10\%$.

\section{DMRG results\label{sec:result}}

We have 
computed ground state properties for ladder systems for various values of $n$%
, $t^{\prime }/t$, and $U/t$ using DMRG. We have kept up to $m=2400$ states
and extrapolated our results to zero truncation errors. As is well known,%
\cite{Bonca-1997} ground state energies (as well as one-point correlation functions%
\cite{White-2007}) can be extracted with great accuracy in this
way. Results have been obtained for system sizes from $2\times 16$
up to $2\times 64$, and then extrapolated to the thermodynamic limit ($%
1/L\rightarrow 0$) using a finite size scaling analysis. From the
extrapolated values, we have extracted $\Delta E _{s}$, $\xi _{s}$, $\Delta
E _{p}$, $\rho _{c}$, and $K_{c}$, as described above.

%
%
%

\begin{figure}[h!]
\includegraphics[angle=0, width=0.4\textwidth]{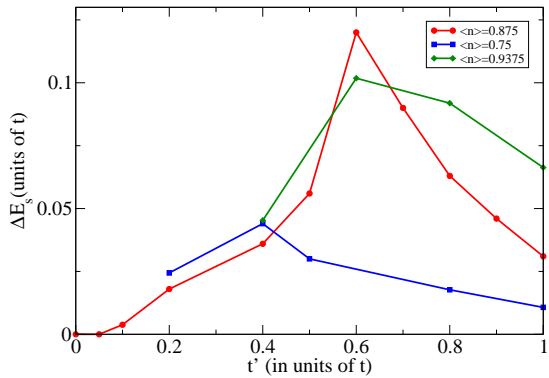}
\caption{ The spin-gap, $\Delta E_s$, of the checkerboard-Hubbard
ladder as a function of $t^\prime/t$ for $n=0.9375, \ 0.875$, and
0.75 at fixed $U=8t$. } \label{fig:gapsn}
\end{figure}

In Fig.\ref{fig:gapsn}, we show $\Delta E _{s}$ for fixed $U/t=8$ as a
function of $t^{\prime }/t$ for
$n=0.9375,\ 0.875$, and $0.75$. 
It has previously been shown that, for the uniform Hubbard two-leg ladder
with $n=0.875$, the spin-gap is maximal for $U=8t$.
\cite{2LegHubbard} Note that the value of $\Delta E _{s}$ rises
from its value for the uniform ladder as $t^{\prime }/t$ is
reduced below $t^{\prime }/t=1$, reaches a maximum value at an
intermediate value of $t^{\prime }/t$, and then drops to zero as
$t^{\prime }/t\rightarrow 0$. For instance, for $n=0.875$, the
maximum value $\Delta E _{s}\approx 0.12t$, which occurs for
$t^{\prime }/t=0.6 $, is approximately 4 times larger than its
value in the uniform ladder. More broadly, we have studied the
spin gap as a function of both $U/t$ and $t^{\prime }/t$; the results for $n=0.875$ are shown in Fig. \ref%
{fig:spin}. One can see that $\Delta E _{s}$ exhibits a broad
maximum for $U$ of order the band-width ($U\sim 4-8t$) and
intermediate inhomogeneity, $ t^{\prime }/t\sim 0.5$. This figure
looks qualitatively similar to the analogous result for the two
dimensional checkerboard Hubbard model obtained previously by
exact diagonalization of a 16 site system in Ref.
\onlinecite{Hong2}; however, in contrast to that study, the
present results are obtained in the thermodynamic limit.

\begin{figure}[h!]
\includegraphics[width=3.4in]{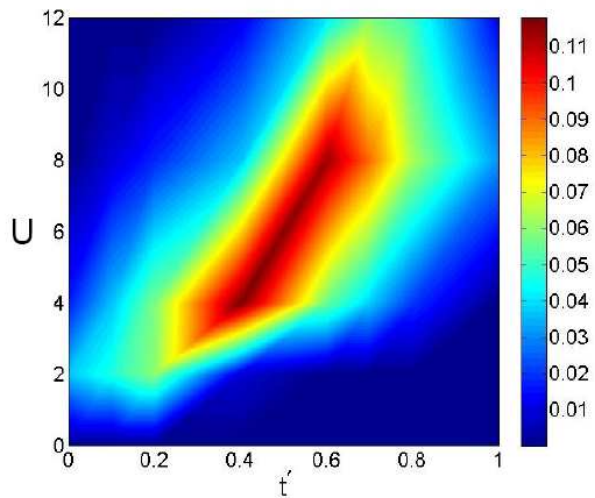}
\caption{$\Delta E_s$ of the checkerboard Hubbard ladder for
$n=0.875$ as a function of $U$ and $t^\prime$, fixing $t=1$.}
\label{fig:spin}
\end{figure}

\begin{figure}[h!]
\includegraphics[width=3.4in]{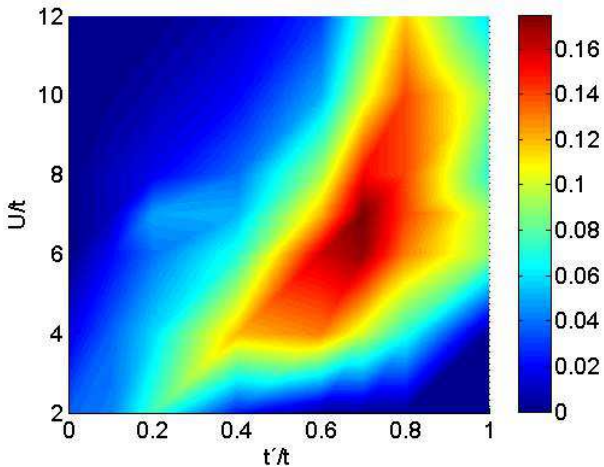}
\caption{The pair binding energy, $\Delta E_p$, of the
checkerboard Hubbard ladder for $n=0.875$ as a function of $U$ and
$t^\prime$, fixing t=1.} \label{fig:pair}
\end{figure}

\begin{figure}[t!]
\includegraphics[width=0.4\textwidth]{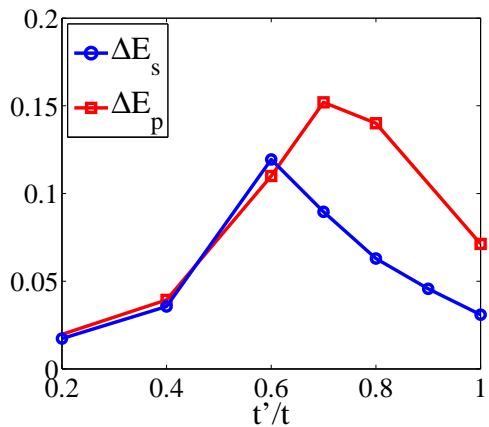}
\caption{$\Delta E_s$ and $\Delta E_p$ for the checkerboard ladder
with fixed $n=0.875$ and $U/t=8$, as a function $t'/t$.}
\label{fig:spin-pair}
\end{figure}

The dependence of $\Delta E _{p}$ on $U/t$ and $t^{\prime }/t$ is
generally similar to that of $\Delta E _{s}$, as can be seen by
comparing the contour plots of these two quantities for $n=0.875$
which are shown in Fig. \ref{fig:pair} and Fig. \ref{fig:spin},
respectively. However, there are interesting differences, as can
be seen in Fig.\ref{fig:spin-pair}, where the two quantities are
plotted as a function of $t^{\prime }/t$ for fixed $U/t=8$ and
$n=0.875$. Note that for $t^{\prime }/t>0.6$, $\Delta E
_{p}>\Delta E_{s}$. This is, presumably, indicative of the
existence of a spin 1 excitonic bound-state for $t^{\prime
}/t>0.6$. A similar result was found in the uniform $t-J$
model\cite{Poilblanc-2000,Poilblanc-2004}.

\begin{figure}[t!]
\includegraphics[width=0.45\textwidth]{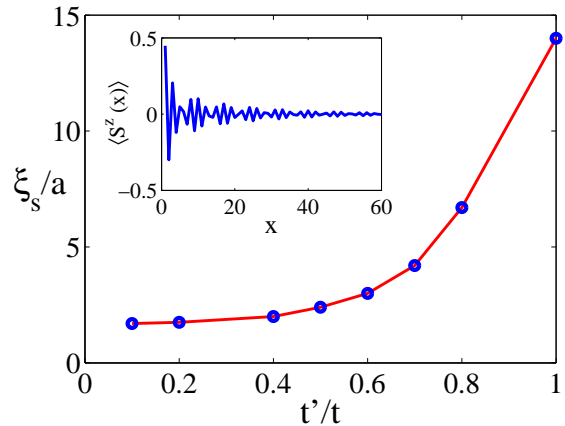}
\caption{The spin correlation length, $\protect\xi_s$, for the
checkerboard ladder with $U=8t$ and $n=0.875$ as a function of
$t^\prime/t$, calculated from Eq. \ref{eq:spincorr}. The inset
shows the expectation value of the spin $\langle S^z \rangle$ for
$U=8t$, $n=0.875$, and $t^\prime/t=1$, as a function of position.
A staggered Zeeman field of strength $t$ has been applied to the
two sites at the left edge of the ladder.} \label{fig:xis}
\end{figure}

In order to calculate $R=\Delta E _{\theta }/\Delta E_s$, we must
compute $\rho _{c}$ and $\xi _{s}$. To obtain $\xi _{s}$, we apply
a relatively strong staggered Zeeman field of magnitude $t$ to the
end sites of the ladder and measure the decay of the staggered
magnetization, $M(j)$ as in Eq. \ref{Mi}. In all cases, we have
found that $M(j)$ decays rapidly on scales short compared to the
length of our longest ladders, so $\xi _{s}$ can be extracted from
the calculations accurately. Representative results for $M(j)$ are
shown in the inset of Fig. \ref{fig:xis}. $\xi _{s}$ as a function
of $t'/t$ is shown in Fig. \ref{fig:xis}, for fixed $n=0.875$ and
$U/t=8$. Note that for $t^{\prime }/t<1/2$, the spin-correlation
length is roughly $3a$, which is of the order of one unit cell of
the checkerboard ladder.

\begin{figure}[t!]
\includegraphics[width=0.45\textwidth]{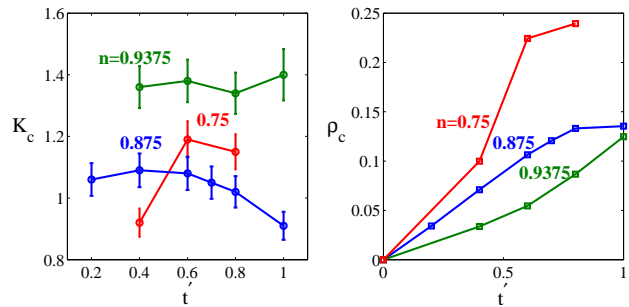}
\vspace{-0.1in}
\caption{Left: the Luttinger parameter $K_c$ as a function of $t^{\prime}/t$ for $%
n=0.75$,$0.875$ and $0.9375$ and $U=8t$. The error bars were
estimated by comparing between the values of $K_c$ obtained from
Eq. \protect\ref{Kc} and Eq. \protect \ref{Kc_density}. Note that
according to our definition of $K_c$, the non-interacting value is
$K_c=2$. Right: The phase stiffness $\protect\rho_c$ (defined in
Eq.\protect\ref{rhoc}) as a function of $t^\prime/t$.}
\label{fig:stiff}
\end{figure}

Next, we calculate both $\rho _{c}$ and $K_{c}$ following the
procedure described above [Eqs.
\ref{compressibility}--\ref{Kc_density}]. The value of $K_{c}$ is
shown in Fig. \ref{fig:stiff} for $n=0.75$, $0.875$, and $0.9375$,
fixing $U/t=8$, as a function of $t^{\prime }/t$. In contrast to
the results for $\Delta E _{s}$ (and somewhat to our surprise),
for $n=0.875$, $K_{c}$ is a weakly varying function of $t^{\prime
}/t$ (and, as it turns out, $U/t$ as well). To a good
approximation, for a wide range of values, we can simply take
$K_{c}\approx 1 $, independent of $t^{\prime }/t$ and $U/t$. Note
that this implies that the superconducting susceptibility diverges
as $T\rightarrow 0$, so that it is reasonable to think of the
ladder as a fluctuating superconductor. (Of course, there is also
a divergent charge-density wave susceptibility, $\chi
_{\mathrm{CDW}}\sim T^{-(2-K_{c})}$, so there is some unavoidable
ambiguity with this simple intuitive picture.) As $n$ is increased
to $0.9375$, $K_c$ increases, consistent with the expectation that
$K_c \rightarrow 2$ as $n \rightarrow 1$. \cite{Schulz-1999}

\begin{table}[h]
\caption{Values of the ratio $R$ defined in Eq. \ref{eq:phase} for
$n=0.875$ and $U=8t$.}
\begin{tabular}{cccccccl}
\cline{1-7}
$t^{\prime }=$ & $0.2$ & $0.4$ & $0.6$ & $0.7$ & $0.8$ & $1.0$ &  \\
\cline{1-7} $R=$ & $3.38$ & $3.06$ & $0.96$ & $1.01$ & $0.99$ &
$0.98$ &  \\ \cline{1-7}
\end{tabular} \label{table:R}
\end{table}

From the measured values of $\xi_{s}$, $\kappa $, and $K_{c}$, the
energy scale characteristic of phase-ordering can be extracted.
Table \ref{table:R} shows the ratio $R$ from Eq. \ref{eq:phase}.
Note that for $t^\prime/t > 0.5$, $R \approx 1$. Thus, at least
crudely, this regime can be thought of as a ``BCS like'' regime,
in which there is a single energy scale, set by $\Delta E_s$,
which characterizes the growth of superconducting correlations.
Depending on precisely what criterion one chooses to quantify the
crossover scale, phase fluctuations will produce a quantitative
difference in the magnitude of the specified scale, but not large
qualitative
effects. 
Therefore, it is reasonable to assert that the values of the
parameters which lead to the largest values of $\Delta E _{s}$
and/or $\Delta E _{p}$ are the \textquotedblleft optimal values
for superconductivity.\textquotedblright

For $t'<0.5t$, we obtain $R\sim 3$, i.e. $\Delta E_\theta > \Delta
E_s$, suggesting that this regime cannot be thought of in terms of
either a naive weak or strong coupling picture. Remarkably, the
transition from $R\sim 1 $ to $R>3$ occurs quite sharply around
$t'=0.5$, close to the point where the spin gap is optimal.

It is interesting to note that for $n=0.75$, $t'/t=0.4$ we find a
sharp decrease of $K_c$ and $\rho_c$. The value of $K_c$ at this
point is smaller than the critical value of 1 at which a static
charge-density wave should be stable\cite{Friedel}, indicating
that this behavior of $K_c$ and $\rho_c$ may be due to a charge-
density wave phase that exists for $n=0.75$, $t'\lesssim 0.4t$.

We thus conclude that for the checkerboard Hubbard ladder, optimal
superconductivity occurs for intermediate values of $U/t\sim 6$,
intermediate inhomogeneity, $t^{\prime }/t\sim 0.6-0.7$, and
electron concentrations near (but not equal to) one electron per
site. We can now ask if this result is special to the checkerboard
pattern, or if it applies more generally to the situation in which
there are multiple sites per unit cell. We thus have repeated
(although not in as much detail) the same calculations for the
dimer ladder (period 1) and the period 3 ladder. (See Fig. \ref
{fig:ladder}.)

\begin{figure}[t!]
\includegraphics[angle=0,width=0.4\textwidth]{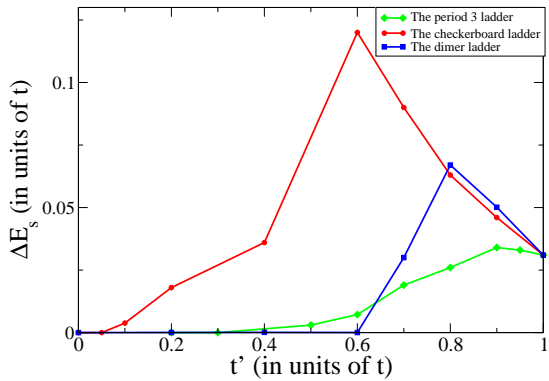}
\caption{ $\Delta E_{s}$ for the three types of inhomogeneous
ladders in Fig. \ref{fig:ladder} is shown at fixed $n=0.875$ and
$U/t=8$ as a function of $t^\prime/t$. The inhomogeneity induced
by breaking up the ladders to period 1, 2 and 3 clusters increases
the spin gap for $t^\prime/t<1$. The increase is most dramatic for
the checkerboard ladder, in which the maximum spin gap is about
$4$ times larger than the spin gap for the uniform ($t'=1$)
system. For the period 1 (dimer) ladder, the enhancement is by a
factor of 2, while for the period 3 ladder the spin gap is only
slightly enhanced (by about $10\%$).} \label{fig:gap2}
\end{figure}

In Fig. \ref{fig:gap2} we exhibit the dependence of the spin-gap of
all three ladders for fixed $U=8t$ and $n=0.875$ as a function of
$t^\prime/t$. In all the cases we see that there is an increase in
the spin gap for some $t^\prime/t<1$.

The result was expected, qualitatively, in the dimer (period one)
case from previous works \cite{2LegHubbard,Noack-1997,Riera-1999},
which found that the spin gap (as well as pairing correlations) is
enhanced upon making $t'$ smaller than $t$ in the dimer (period 2)
ladder. In the case of the period three ladders, there is a very
weak increase of the spin gap, which occurs at $t^\prime/t \sim
0.9$.

In Refs. \onlinecite{Noack-1997}, it was argued that the
enhancement of superconducting correlations in the dimer ladder is
due to the increase in the density of states close to the ``Van
Hove'' point, in which one of the two bands of the two--leg ladder
becomes unoccupied. Beyond this point, there is only a single band
crossing the Fermi level, and the system is likely to behave as a
single--component Luttinger liquid. Therefore the superconducting
signatures are rapidly suppressed. Consistently with this picture,
in the dimer ladder, we find that the spin gap collapses to zero
below $t'/t \lesssim 0.6$. In the period 2 (the checkerboard
ladder) and period 3 cases, however, no such sudden suppression of
the spin gap is observed as $t'/t$ is reduced below the optimal
point. This leads us to believe that the mechanism of the
enhancement of the spin gap for $t'<t$ in the checkerboard and
period 3 ladders is unlikely to be related to a proximity to a Van
Hove point.

Note also that for all the inhomogeneous patterns in Fig.
\ref{fig:gap2}, the spin gap seems to reach zero at a critical
$t'_c>0$ (which is different for each pattern). In particular, for
the ``checkerboard'' pattern, $t'_c\sim 0.05t$. It is likely that
for $t'<t'_c$, the Luther-Emery phase gives way to a Luttinger
liquid phase with one gapless charge mode and and gapless spin
mode (or more), although more work is needed to establish that.

Overall, among all the patterns we have reported, the optimal
ladder for
superconductivity is a checkerboard ladder with $U=6t$, $t^\prime/t=0.6-0.8$%
, and $n=0.875$, for which $\Delta E_s=0.12t$, $\Delta E_p =0.16t$.


\section{Extension to Quasi 1D\label{sec:mean}}

\begin{figure}[t!]
\includegraphics[width=0.4\textwidth]{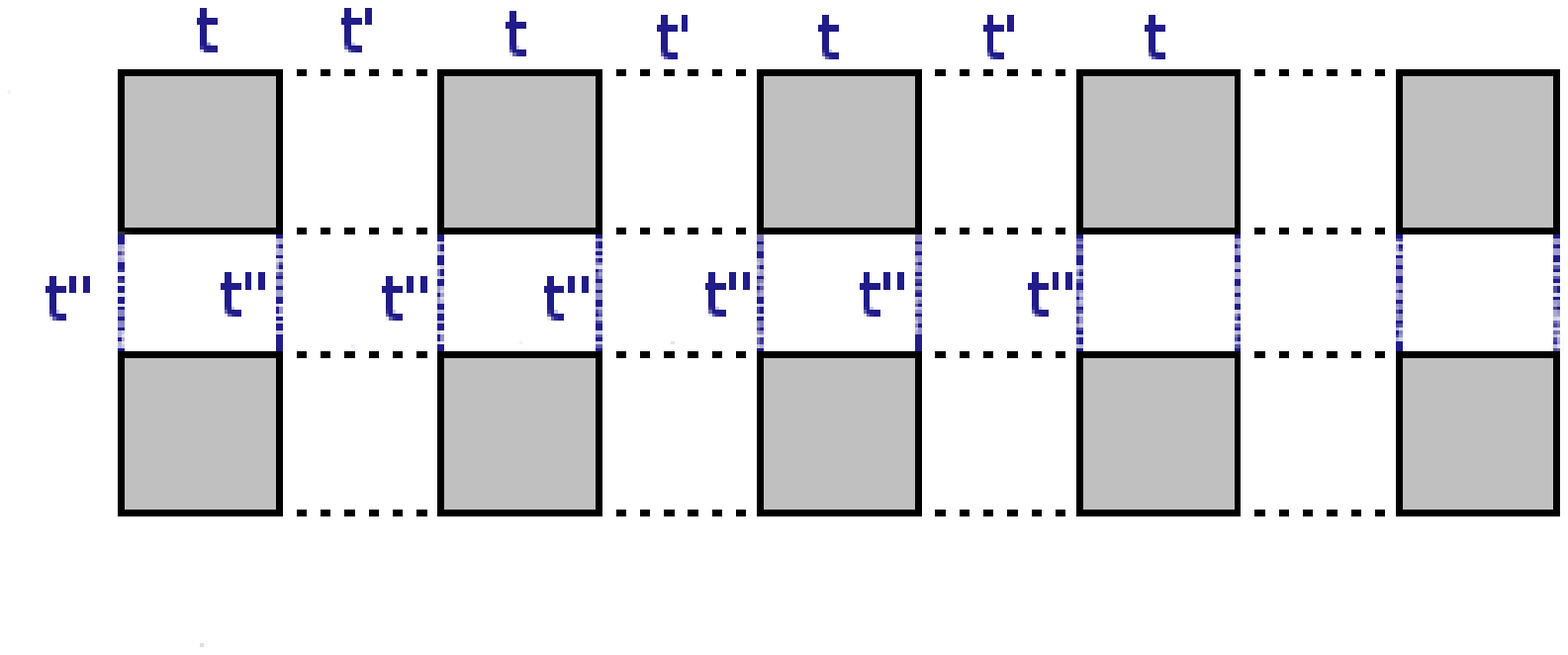}
\caption{A system of coupled checkerboard ladders, connected with by a
single particle tunneling matrix element $t^{\prime \prime }$}
\label{fig:inter}
\end{figure}

Above, we have argued that the superconducting tendency in the 
checkerboard-Hubbard ladder is optimized for an intermediate value of $%
t^{\prime }/t$. However, since the superconducting $T_{c}$ of that system
(as in any one-dimensional system) is strictly zero, one can worry that this
statement may depend on how one chooses to measure the strength of the
superconducting correlations. We will now 
consider a system composed of an array of parallel 
checkerboard-Hubbard ladders 
coupled weakly in the direction transverse to the ladders, in which $T_{c}$
can be estimated in a controlled way based on the solution of the
single-ladder problem. We will show that $T_{c}$ is maximal for $\frac{%
t^{\prime }}{t}<1$. Thus, in this system, $T_{c}$ is indeed optimized when
the electronic structure is non-uniform; \emph{i.e.}, there is an
\textquotedblleft optimal degree of inhomogeneity\textquotedblright\ for
superconductivity.

The quasi-1D system of coupled 
checkerboard Hubbard ladders is depicted in Fig. \ref{fig:inter}. The
ladders are coupled by a single particle tunneling matrix element $t^{\prime
\prime }$. We fix the value of $t^{\prime \prime }\ll t,t^{\prime }$, and
estimate $T_{c}\left( t^{\prime }/t\right) $ from an inter-chain mean field
theory, described in Appendix A. From the numerical results for the
checkerboard-Hubbard ladder with $n=0.875$ and $U=8t$ we recall
that $K_{c}\left( \frac{t^{\prime }}{t}\right) $ $\approx 1$ over
the entire range $0<t^{\prime }\leq 1$ (see Fig. \ref{fig:stiff}).
We therefore fix $K_{c}=1$, independent of $t^{\prime }$. The
resulting expression for $T_{c}$ is
\begin{equation}
T_{c}\sim \frac{K(\sqrt{1-x^2})}{x} 
\Delta E _{s}\left( \frac{%
at^{\prime \prime }}{v_{c}}\right) ^{2}\text{.}  \label{Tc}
\end{equation}%
Here, $x\equiv v_s/v_c$, and $K\left(x\right) $ is a complete elliptic integral of the first kind. Note that $%
T_{c}$ depends on $t^{\prime }/t$ through $v_{s}$, $v_{c}$ and
$\Delta E _{s} $. As $t^{\prime }$ decreases, both $v_{c}$ and
$v_{s}$ decrease; their ratio, however, is found to be
approximately constant as a function of $t'/t$ down to about
$t'/t=0.5$. ($v_s$ is obtained by using the estimate $\Delta E_s
\xi_s$, where both $\Delta E_s$ and $\xi_s$ are calculated from
DMRG.)
$\Delta E _{s}\left( \frac{t^{\prime }}{t}\right) $, on the other hand,
has a maximum for $\frac{t^{\prime }}{t}<1$. Therefore, as $\frac{t^{\prime }%
}{t}$ is reduced from $1$, $T_{c}\left( t^{\prime }\right) $ necessarily
increases, and reaches a maximum for some $t_{\max }^{\prime }<t$.

\section{Discussion\label{sec:concl}}

The present study, along with a variety of other recent
studies\cite{Hong,Hong2,Baruch-2010,Maier-2010}, provide strong
support for a number of intuitively appealing ideas concerning the
physics of the superconducting $T_c$ in unconventional
superconductors in which the pairing arises directly from the
repulsive interactions between electrons: 1) The highest
superconducting transition temperatures occur at intermediate
interaction strength, when $U$ is comparable to the band-width. (A
corollary of this is that materials which are studied because of
their high transition temperatures are also likely to exhibit more
general signatures of lying in an intermediate coupling regime;
here, theoretical results from both weak and strong coupling
approaches must be extrapolated, at best, to the limits of their
regimes of applicability.) 2) Certain mesoscale structures
(``optimal inhomogeneity''\cite{Steve&Fradkin}) can lead to
enhanced superconducting pairing, although clearly if the system
is too strongly inhomogeneous, that always leads to a suppression
of global phase coherence. 3) While short-range magnetic
correlations, possibly of the sort envisioned in the putative RVB
state of a quantum antiferromagnet or in certain theories of a
spin-fluctuation exchange mechanism, may well be important for
pairing, longer range magnetic correlations, especially of the
sort one would expect near a magnetic quantum critical point, do
\emph{not} appear to be particularly favorable for
superconductivity. (This final conclusion follows from a
comparison of the $t^\prime/t$ dependence of the
magnetic correlation length and the superconducting pairing in Figs. \ref%
{fig:xis} and \ref{fig:pair}, respectively.)

In addition, we found that the two-leg ladder at intermediate
coupling (with $U$ of the order of the bandwidth) and close to
half filling is, in many respects, surprisingly well described as
a ``BCS--like'' superconductor, in which there is a single
crossover energy scale from the ``normal'' to the
``superconducting'' state (rather than two separate scales,
associated with pairing and phase coherence). This is based on the
fact that the ratio of the pairing and phase coherence scales
(defined in Eq. \ref{eq:phase}) is close to its weak-coupling
value, which justifies the identification of the spin gap $\Delta
E_s$ (or the pair binding energy $\Delta E_p$) as the relevant
energy scale for superconductivity.


Finally, there are a couple of unresolved issues and further
directions we would like to highlight: 1) The extrapolation of the
present results to higher dimensions is, of course, the most
important open issue. The strong qualitative similarity between
the present results and those obtained by exact diagonalization
and CORE calculations on relatively small 2D clusters certainly
encourages us to believe that the results obtained here give
insight into the behavior of the higher dimensional problem. In
this context, it might be useful to carry out similar calculations
on 4 leg and possibly even 6 leg ladders and cylinders, although
it is considerably more difficult to extend these results to such
long systems as are accessible for the 2 leg ladder. 2) It is not
clear exactly what aspects of the local electronic structure are
essential features of an optimal inhomogeneity for
superconductivity. In the present case, it is notable that
pair-binding does not occur on an isolated dimer or six-site
rectangle for any value of $U/t$, while there is pair-biding on an
isolated square for $U/t < 4.6$. However, this observation does
not provide an entirely satisfactory account of our findings,
since the optimal pairing in the checkerboard ladder occurs for $
U/t = 4-8t$, where the pair-binding energy of an isolated square
is either small or negative.

\begin{acknowledgements}
We thank Ehud Altman, Assa Auerbach, Malcolm Beasley, Thierry
Giamarchi, Lilach Goren, Dror Orgad, Didier Poilblanc, Doug J.
Scalapino, Alexei Tsvelik, and Wei-Feng Tsai for useful
discussions. This research was supported by the NSF under Grants
DMR-0531196 (SAK and GK), DMR-0907500 (SRW), DMR-0705472 and
DMR-0757145 (EB).
\end{acknowledgements}

\section{Appendix: Inter-chain mean--field theory}

In this Appendix, we describe the inter--chain mean--field
treatment of the quasi one dimensional system described in Sec.
\ref{sec:mean}. This procedure is quite
standard\cite{Orignac-1997,Giamarchi-1999,Carlson-2000}. We
consider an array of plaquette ladders, modelled by Luther--Emery
liquids. For simplicity, we will assume that each ladder is a
single component system with a spin gap $\Delta E _{s}$. (The
extension to the case of a two-component system is
straightforward, and the result is qualitatively the same.) The
ladders are coupled by an inter-chain hopping term of the form:
\begin{equation}
H_{\perp }=-t^{\prime \prime }\sum_{\sigma ,P=\pm }\sum_{n}\int dx\psi
_{P\sigma }^{\dagger }\left( x,n\right) \psi _{P\sigma }\left( x,n+1\right)
\text{,}
\end{equation}%
where $\psi _{P\sigma }^{\dagger }\left( x,n\right) $ ($P=\pm $)
creates a right or left moving electron with spin $\sigma
=\uparrow ,\downarrow $ at position $x$ in chain $n$. Next, we
integrate out degrees of freedom of lengthscales smaller than the
spin correlation length $\xi _{s}\sim \frac{v_{s}}{\Delta E
_{s}}$. Over such lengthscales, the system is essentially gapless
and can be treated as a Luttinger liquid. To second order in $
t^{\prime \prime }$, the following effective inter--chain action
is generated:
\begin{widetext}
\begin{equation}
S_{\perp }^{\mathrm{eff}}=(t^{\prime \prime })^{2}\sum_{\sigma \sigma
^{\prime },n}\int dxd\tau \int dx^{\prime }d\tau ^{\prime }\langle \mathcal{T%
}\psi _{+,\sigma }^{\dagger }\left( x,\tau ,n\right) \psi _{+,\sigma }\left(
x,\tau ,n+1\right) \psi _{-,\sigma ^{\prime }}^{\dagger }\left( x^{\prime
},\tau ^{\prime },n\right) \psi _{-,\sigma ^{\prime }}\left( x^{\prime
},\tau ^{\prime },n+1\right) \rangle _{0,>}\text{,}  \label{Sperp}
\end{equation}
\end{widetext}
where $\langle \dots \rangle _{0,>}$ denotes averaging over the
\textquotedblleft fast\textquotedblright\ (short-wavelength) degrees of
freedom [using the decoupled ($t^{\prime \prime }=0$) action], and $\mathcal{%
T}$ denotes time ordering. Since we are essentially performing a
\textquotedblleft coarse graining\textquotedblright\ step, increasing the
cutoff of the theory from the lattice constant $a$ to $\xi _{s}$, the region
of integration in Eq. \ref{Sperp} is $\sqrt{\left( x-x^{\prime }\right)
^{2}+v_{s}^{2}\left( \tau -\tau ^{\prime }\right) ^{2}}<\xi _{s}$. In order
to evaluate the integrand, we write the fermionic fields in bosonized form: $%
\psi _{P\sigma }\sim e^{i\sqrt{\pi }\left( \theta _{\sigma }+P\varphi
_{\sigma }\right) }$, where $\varphi _{\sigma }$ and $\theta _{\sigma }$ are
dual bosonic fields which satisfy $\left[ \varphi _{\sigma }\left( x\right)
,\partial _{x}\theta _{\sigma ^{\prime }}\left( x^{\prime }\right) \right]
=i\delta _{\sigma \sigma ^{\prime }}\delta \left( x-x^{\prime }\right) $. As
usual, we introduce also charge and spin fields defined as $\varphi
_{c,s}=\left( \varphi _{\uparrow }\pm \varphi _{\downarrow }\right) /\sqrt{2}
$, and similarly $\theta _{c,s}=\left( \theta _{\uparrow }\pm \theta
_{\downarrow }\right) /\sqrt{2}$. We define the fermionic Green's function $%
\mathcal{G}\left( x,\tau \right) =\langle \mathcal{T}\psi _{+,\uparrow
}\left( x,\tau ,n\right) \psi _{-,\downarrow }\left( 0,0,n\right) \rangle
_{0,>}$. Expressing $\mathcal{G}\left( x,\tau \right) $ in terms of the
bosonic fields,
\begin{widetext}
\begin{equation}
\mathcal{G}\left( x,\tau \right) \sim \left\langle e^{i\sqrt{2\pi }\left[
\frac{\theta _{c}+\theta _{c}^{\prime }}{2}+\frac{\theta _{s}-\theta
_{s}^{\prime }}{2}+\frac{\varphi _{c}-\varphi _{c}^{\prime }}{2}+\frac{%
\varphi _{s}+\varphi _{s}^{\prime }}{2}\right] }\right\rangle _{0,>}\sim
\left\vert \frac{a^{2}}{x^{2}+\left( v_{c}\tau \right) ^{2}}\right\vert ^{%
\frac{1}{8K_{c}}}\left\vert \frac{a^{2}}{x^{2}+\left( v_{c}\tau \right) ^{2}}%
\right\vert ^{\frac{K_{c}}{8}}\left\vert \frac{a^{2}}{x^{2}+\left( v_{s}\tau
\right) ^{2}}\right\vert ^{\frac{1}{4}}e^{i\sqrt{2\pi }\left[ \bar{\theta}%
_{c}+\bar{\varphi}_{s}\right] }\text{,}  \label{G}
\end{equation}
\end{widetext}
where we have used the shorthand notation $\theta _{c}\equiv \theta
_{c}\left( x,\tau \right) $, $\theta _{c}^{\prime }\equiv \theta _{c}\left(
x^{\prime },\tau ^{\prime }\right) $, $\bar{\theta}_{c}\equiv \theta
_{c}\left( x/2,\tau /2\right) $, and similarly for $\theta _{s}$, $\varphi
_{c}$ and $\varphi _{s}$. Plugging $\mathcal{G}\left( x,\tau \right) $ into
Eq. \ref{Sperp} and performing the integral, we get that the following
inter--chain Josephson coupling term:
\begin{equation}
H_{\perp }^{\mathrm{eff}}=-J_{\perp }\sum_{n}\int dx\Phi \left( x,n\right)
\Phi ^{\dagger }\left( x,n+1\right) +\mathrm{H.c.}\text{,}
\end{equation}%
where $\Phi \left( x,n\right) =\psi _{R\uparrow }\psi _{L\downarrow }-\psi
_{R\downarrow }\psi _{L\uparrow }\sim e^{i\sqrt{2\pi }\theta _{c}}\cos \sqrt{%
2\pi }\varphi _{s}$ and
\begin{equation}
J_{\perp }\sim K\left( \sqrt{1-\left( \frac{v_{s}}{v_{c}}\right) ^{2}}%
\right) \left( \frac{at^{\prime \prime }}{v_{c}}\right) ^{2}\frac{v_{c}}{a}%
\text{.}  \label{Jperp}
\end{equation}%
Here, $K\left( \alpha \right) =\int_{0}^{\pi /2}d\lambda
/\sqrt{1-\alpha ^{2}\sin ^{2}\lambda }$ is a complete elliptic
integral of the first kind. Eq. \ref{Jperp} contains a
$K_c$--dependent prefactor, which we omit.

The mean--field equation for $T_{c}$ is
\begin{equation}
zJ_{\perp }\chi \left( T_{c}\right) =1\text{,}  \label{MF}
\end{equation}%
where $\chi \left( T\right) $ is the superconducting susceptibility of a
single chain, and $z$ is the number of nearest--neighbor chains (\emph{e.g.}%
, for a two dimensional array of checkerboard ladders, $z=2$).
Inserting Eqs. \ref{chi},\ref{Jperp} in the mean--field equation
(\ref{MF}), and using the fact that for the checkerboard Hubbard
ladder $K_{c}\approx 1$ over a wide range of parameters, we obtain
Eq. \ref{Tc} for $T_{c}$.

\bibliography{bibliography}

\end{document}